# High redshift radio galaxies[1]

R. A. E. Fosbury[2]

Space Telescope – European Coordinating Facility
Karl-Schwarzschild-Straße 2
D-85748 Garching bei München, Germany

**ABSTRACT**

There is considerable evidence that powerful radio quasars and radio galaxies are orientation-dependent manifestations of the same parent population: massive spheroids containing correspondingly massive black holes. Following the recognition of this unification, research is directed to the task of elucidating the structure and composition of the active nuclei and their hosts to understand the formation and evolution of what we expect to become the most massive of galaxies. In contrast to the quasars, where the nucleus can outshine the galaxy at optical/near infrared wavelengths by a large factor, the radio galaxies contain a 'built-in coronograph' that obscures our direct view to the nucleus. These objects present our best opportunity to study the host galaxy in detail. Of particular interest are those sources with redshifts greater than about 2 that represent an epoch when nuclear activity was much more common that it is now and when we believe these objects were in the process of assembly. In combination with high resolution imaging from space (HST), optical spectropolarimetry with Keck II allows us to clearly separate the scattered nuclear radiation from the stellar and gaseous emission from the host galaxy. The rest-frame ultraviolet emission line spectra suggest that rapid chemical evolution is occurring at this epoch. Near infrared spectroscopy with the VLT is giving us access to both the lines and continuum in the rest-frame optical spectrum, allowing investigations of the evolved stellar population and extending the composition analysis with measurements of the familiar forbidden-line spectrum.

Keywords: Radio galaxy, quasar, galaxy evolution, spectropolarimetry, IR-spectroscopy, scattering, natural coronograph.

## 1. INTRODUCTION

The ability of the Hubble Space Telescope (HST) to obtain very deep, exquisite resolution images of the sky[1,2] has kindled a lively observational interest in the evolution of galaxies early in the history of the Universe. The new, large, groundbased telescopes provide the multicolour photometric and spectroscopic power to make this a mature field of astrophysics. Techniques of colour-selection, most notably the use of the Lyman continuum and Lyman forest breaks produced by neutral hydrogen absorption[3], have made it possible to sift the few high redshift objects from the multitude of fainter, closer galaxies. The identification of radio sources, however, remains an efficient method of selecting very distant galaxies and quasars. In particular, the use of radio spectral index selection has provided many of the radio galaxy identifications at the highest redshifts[4]. The radio sources are important because they appear to mark the sites of formation of the most massive galaxies at the cores of assembling clusters.

The radio galaxies occupy a special place in the study of early galaxy formation since, although they are believed to contain an active quasar core, our direct line of sight to the nucleus is obscured by very optically thick — at least from UV to near-IR wavelengths — material in its close (<1pc) vicinity. This removes much of the glare associated with the study of the host galaxies of quasars and allows us to see remaining galaxian components more clearly. The presence of the quasar, however, has a significant effect on the surrounding galaxy, particularly on the state of the gas which sees the hard radiation field directly. These interactions need to be understood if we are to extract the story of massive galaxy formation and evolution from the line and continuum radiation we observe.

This paper describes projects underway which are designed to identify the various components which contribute to the radiated spectrum of the host galaxies. The emphasis here is on the redshifted UV and optical spectrum where the dominant players are starlight, scattered quasar light and gaseous line emission excited by the quasar radiation field. An important part of the story will be told at much longer wavelengths, however, where dust is re-radiating much of the energy derived from the stars and active nucleus.

---

[2] rfosbury@eso.org, www.stecf.org/~rfosbury. Affiliated to the Astrophysics Division of the Space Science Department, European Space Agency.

## 2. THE OBJECTS

Radio galaxies and quasars are now found at redshifts beyond 5 and are so the most distant objects clearly identified before we reach the source of the cosmic microwave background radiation. At our present epoch they are relatively rare and the objects which were so common at redshifts beyond $z \sim 2$ are poorly represented in our local Universe. The redshift range between 2 and 3 or so represents the epoch when nuclear activity was common and is likely to be the time when the massive spheroids, destined to become giant elliptical galaxies, were being assembled. This redshift range also has the practical advantage of moving the ultraviolet spectrum, longwards of the Ly (1215Å) line, into the optical spectrum accessible to groundbased telescopes. This spectral region contains several strong resonance lines from abundant elements as well as a host of other lines of diagnostic value. It is also the place where strong photospheric lines might be seen from young stars recently formed in the galaxy and it contains the classical dust extinction signature near 2175Å.. The familiar optical spectrum, containing the forbidden and recombination lines used for diagnosing the state of the interstellar medium (ISM) and also the maximum continuum emission from an evolved stellar population, is shifted into the near-IR region where the new, large groundbased telescope infrared spectrometers can observe.

Most of the radio galaxies known at these redshifts have been imaged with HST at optical[5] and near-IR[6] wavelengths. These images show rather clumpy structures suggesting the assembly of galaxies from smaller components and possibly also the presence of dust obscuration. Only at the lower redshifts and in the near-IR do the objects start to resemble present-day ellipticals. The images are generally elongated and aligned with the axes of their radio sources, a phenomenon first described in 1987 and called the 'alignment effect' [7,8].

## 3. OPTICAL SPECTROPOLARIMETRY

The technique of polarimetry is used here to separate scattered quasar light from starlight in the host galaxy. It has been known for some years[9] that the blue/UV continuum from radio galaxies is often linearly polarized with an electric-vector orientation consistent with the hypothesis that all or part of the alignment effect could result from the scattering of anisotropically emitted radiation from the nucleus shining into broad bi-conical regions aligned with the radio axis. This early polarimetry with 4m telescopes was mostly carried out in imaging-mode since the faintness of the sources prohibited the further spreading of the light. Spectropolarimetry provides a more powerful diagnostic of the scattering mechanisms involved, associated with free electrons or dust, and became possible to carry out with the polarimetry module LRISp[10] on the Keck LRIS spectrograph[11]. The high photon counts necessary to perform continuum polarimetry to a precision of a few % per spectral resolution element implies observing times, even with a 10m telescope, of several hours. A by-product of this polarimetry, however, is spectrophotometry of good s/n which allows measurement of many emission and absorption lines in addition to the continuum energy distribution. The first results, for two sources, of this programme are published[12] and the full description is in Vernet et al. (in prep.). Twelve sources have been observed so far with redshifts $2.3 < z < 3.6$ with the data reduced to the $I$, $Q$ and $U$ Stokes parameters and thence to fractional linear polarization $P$ and PA of the electric vector . Error estimates are derived from Monte-Carlo simulations of the observing process since photon statistics is the major source of uncertainty. The slit is aligned with the radio axis of the source and the data are corrected for extinction in our Galaxy (only). Since the data are not always obtained in photometric conditions, the fluxes are scaled to HST magnitudes where available.

## 4. IR SPECTROSCOPY

While measurements of optical emission lines in radio galaxies in this redshift range[13] have been made in order to remove the inherent ambiguities in the interpretation of broad-band infrared photometry, the detection of the continuum at these wavelengths is difficult. The presence of ISAAC[14] on the VLT and NIRSPEC[15] on Keck has now made such tasks possible. With redshifts $2.1 < z < 2.7$, the major emission lines ([OII] 3727Å; H and [OIII] 5007Å; H ) are conveniently placed in the J, H and K atmospheric windows respectively. Our programme with ISAAC is being carried out to provide such IR spectroscopy for about half of the sources which have already been observed at shorter wavelengths with LRISp. This combination gives us almost complete wavelength coverage, with a spectral resolving power R = / of several hundred, from Ly to H . Six sources have so far been observed with ISAAC in the H and K windows with three having J-band observations as well. Typical observing times are one hour per band although good continuum measurements really demand somewhat longer.

## 5. RESULTS

A typical result of a spectropolarimetric observation is shown in Fig. 1 where the panels show total flux — with two scalings to show lines and continuum, $P$ and . TXS0828+193 is a source with an intermediate continuum polarization of around 8% and a rich emission line spectrum. The $P$ panel shows measurements in continuum and (narrow) emission line bands. The latter have a lower fractional polarization because the bulk of the narrow line photons reach us directly from the extended, ionized ISM of the galaxy. Residual polarization in these bands is due to scattered continuum and broad lines from the active nucleus. The sources with stronger polarization show clear broad wings on the strong permitted UV lines of about the strength and width expected from the hidden quasar. The value of $P$, measured in the continuum just longward of Lyα, ranges from <3% to over 20% and is either flat or rising slightly to the blue. The E-vector is always perpendicular to the extension of the UV structure seen at HST resolution.

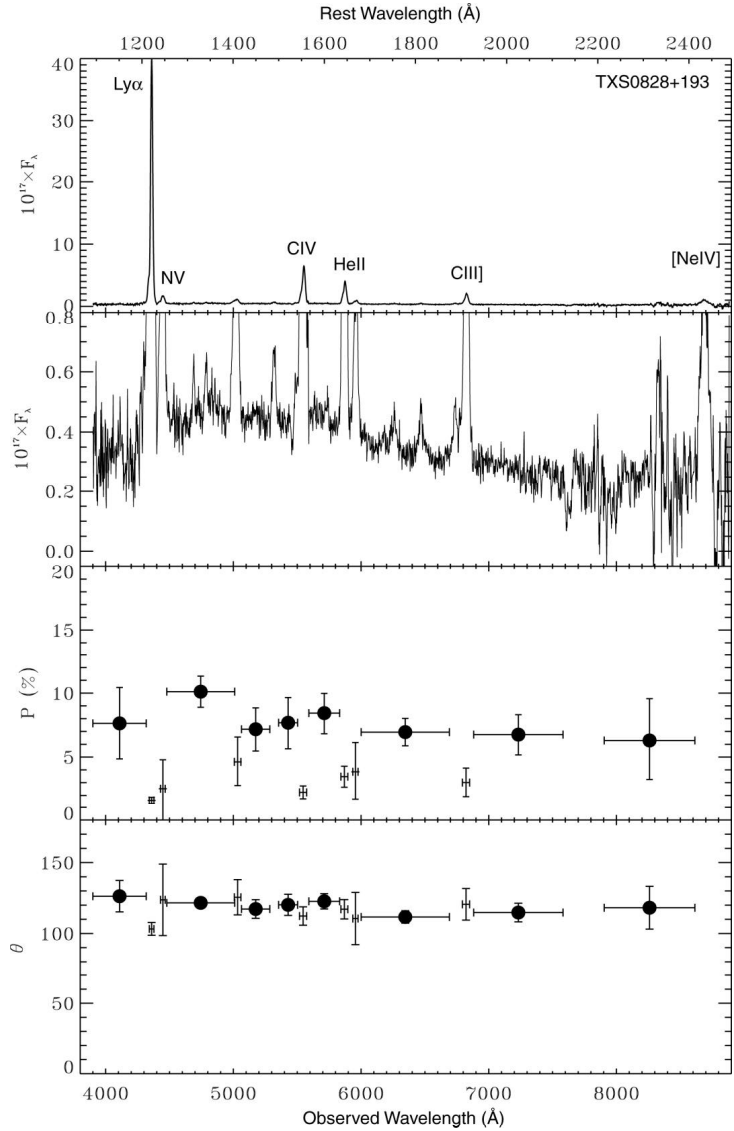

**Figure 1.** Spectropolarimetry of the z = 2.572 radio galaxy TXS0828+193 obtained with LRISp on the Keck II telescope. The top two panels show the observed flux per unit wavelength in units of $10^{-17}$ erg cm$^{-2}$ s$^{-1}$ Å$^{-1}$. The third panel shows the fractional polarization in % and the fourth, the position angle on the sky (N through E) of the electric-vector. The filled circles represent continuum bands while the crosses indicate measurements made across strong emission lines. The increase in noise towards the red is due to the strong night-sky emission in this region. The error-bars are 1σ, computed from Monte-Carlo simulations of the observations.

In all the sources, both emission lines and continuum are spatially extended at least by a few arcseconds with the Lyα emission extending up to 20" in some cases. The emission lines show complex kinematic structures extending up to ± 2,000 km s$^{-1}$.

An unexpected result is the similarity of the continuum colours between 1200 and 2500Å in all the sources. The peak of $F_\lambda$ near Lyα and the minimum near 2200Å is typical with the slope showing little variation in this range.

Although interstellar absorption lines are seen, these are weaker than is usual in Lyman-break galaxies at similar redshifts[16]. We see no convincing evidence for photospheric absorptions from OB stars but our spectral resolution makes our observations insensitive to these weak lines. The SiII and CII resonance lines behave in a complex manner appearing sometimes in emission and sometimes in absorption.

The emission line ratios indicate a rather uniform state of ionization of the gas. There is little variation in the CIII]/CIV and the HeII/CIV ratios but much larger variations in Lyα/CIV and NV/CIV.

Amongst the spatially integrated properties of the sample of sources measured with LRISp, there are two outstanding correlations:

1. The strength of the Lyα emission line (normalised to the strength of CIV) decreases as the continuum polarization increases.
2. The NV/CIV ratio is positively correlated with the continuum polarization.

The infrared data have not yet been fully analysed but show a large variation in the behaviour of the continuum and broad Hα emission line suggesting that direct, reddened, quasar light can be seen in the optical spectrum of some of the sources. This would have caused them to be classified as broad line radio galaxies at lower redshift. A composite spectral energy distribution for MRC1138-262 is shown in Fig. 2, illustrating one of the sources which shows significant long wavelength quasar continuum and broad Hα.

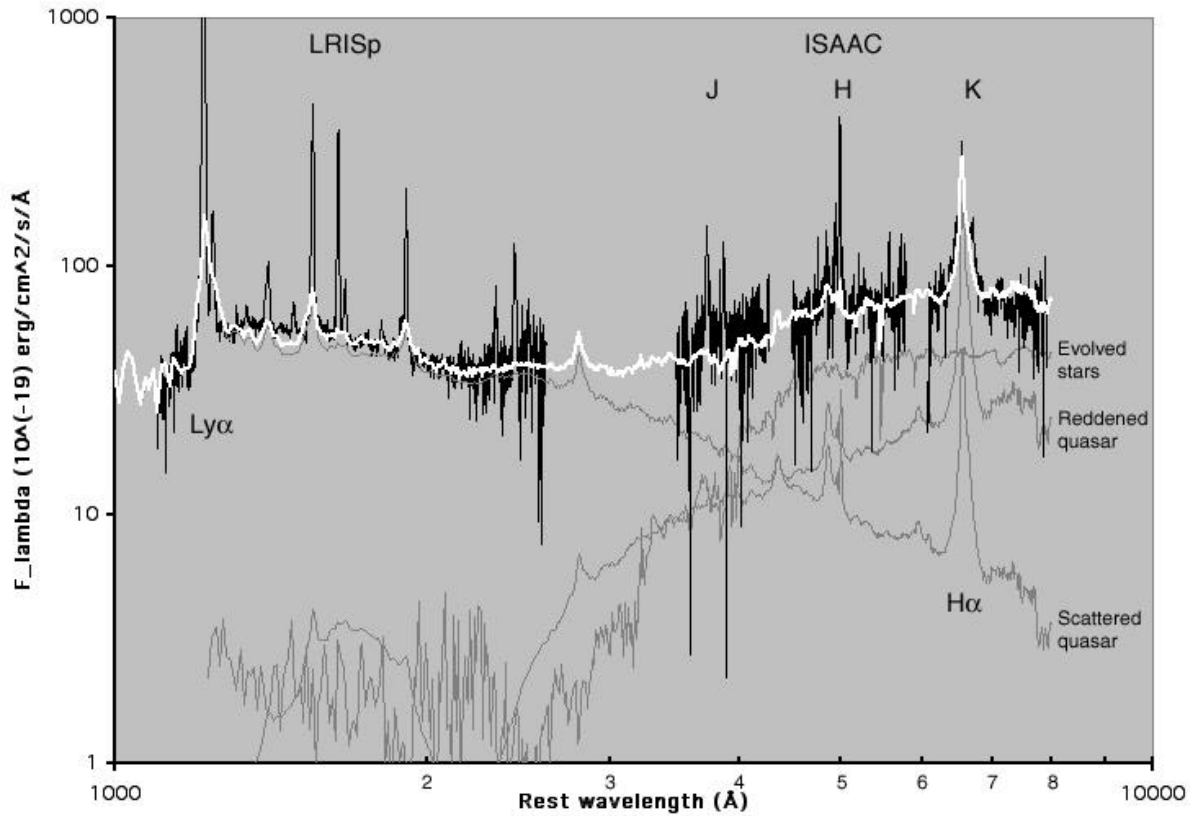

**Figure 2.** Schematic composite spectrum of a radio galaxy. The blue (LRISp) data (black) are an average of 6 observed sources scaled to the known[17] UV continuum flux of MRC1138-262. The red data (black) are ISAAC measurements of MRC1138-262. The three grey lines indicate the components used to model the energy distribution: a scattered quasar (see text); the same quasar seen directly through an extinction screen of Galactic-type dust; and an evolved stellar population. The white line is the sum of these three. The elliptical galaxy used here is probably too red to represent an evolved population at this epoch, as indicated by the poor fit in the J-band. Note that there is no attempt here to fit the narrow emission lines which arise predominantly in the extended ISM and are seen directly.

## 6. DISCUSSION

The principal areas of analysis concern, firstly, the nature of the continuum radiation and, secondly, the interpretation of the emission line spectrum in terms of the properties of the ISM. The absorption spectra reveal some intriguing characteristics but the current observations are not well-suited to their detailed study, a higher spectral resolving power being required.

The existence of significant linear polarization of the restframe UV spectra of all but one of our sources indicates that scattered quasar light is contributing to the observed continuum. The contribution of nebular continuum in the UV, computed from the strength of the H and He recombination lines, is generally less than 20% of the observed continuum. Optically thin scattering models, using a geometry based on detailed observations of low redshift sources, show that the expected polarizations are significantly diluted by the necessity to integrate over a large range of scattering angles implied by the bi-conical geometry[18]. These dust scattering models reproduce the magnitudes of the fractional polarizations observed in radio galaxies but they need to introduce some subsequent dust extinction in order to reproduce the observed energy distributions. Our approach in this study is to note that the distribution of dust in the host galaxies is likely to be very clumpy with a wide range of dust optical depths being available to an escaping photon. Under these circumstances, a natural 'luminosity weighting' process operates with the emergent flux being dominated by photons experiencing an optical depth close to unity[19, 20]. These models also show that the fraction of the flux emerging from such a clumpy geometry is only very weakly dependent on the radial optical depth from the photon source to the edge of the scattering medium. Such scattering is naturally grey over a wide wavelength range regardless of the details of the wavelength dependence of the scattering and absorption cross-sections. Spectral signatures of the scattering/absorption process are only imprinted on the emergent spectrum when the two cross-sections have different wavelength dependencies as they do over the 2200Å graphite feature in Galactic dust[21].

We have therefore used a very simple model to represent the scattered UV light in the radio galaxies, assuming that the emergent flux, $F_{RG}$, is given by:

$$F_{RG} \sim \gamma F_{quasar} \sigma_{scat} \exp(-\tau_{ext})$$

Where $\gamma$ is a geometrical factor. Given that the scattered flux at any wavelength comes from $\tau_{ext} \sim 1$, the $F_{RG}$ becomes proportional to the ratio of the scattering to the extinction cross-sections. This is shown in Fig. 2 where a template quasar has been multiplied by this ratio (for MRN dust) to produce a remarkably good fit to the average radio galaxy spectrum in the UV. The only free-parameter in this fit is the value of $\gamma$ which is a few percent if the radio galaxies host luminous quasars.

Given the excellent match between the scattered quasar and the radio galaxy energy distributions in the UV, there are quite severe constraints on the colour of any blue stellar population which might be present. For the sources with significant polarization, the UV spectrum is dominated by scattered light with starlight contributing less than about 20%.

In the restframe optical spectrum, the scattered continuum is still present — and we expect some scattered broad H$\alpha$ — but two other contributors can easily dominate it: an evolved (red) stellar population which has long been suggested by the behaviour of the K-band Hubble diagram for radio galaxies[22], and direct, reddened quasar light in some cases. Determining the nature of an evolved population will require a careful subtraction of the direct and scattered quasar components in individual cases and will probably rely on the measurement of substantial spectral features like the 4000Å-break.

The emission lines from the global ISM of the host galaxy come predominantly from those zones that see the ionizing radiation from the quasar directly. While a number of mechanisms may be responsible for heating and ionizing this gas, including shocks and cosmic rays, the ionization of the warm (~ $10^4$K) ISM in the hosts of these powerful radio galaxies is thought to be dominated by the hard AGN radiation field. The presence of this galaxy-scale gas, illuminated by an intense ionizing radiation field resulting in bright emission lines spanning a broad range of ionization states, presents an opportunity for detailed studies of its physical state and chemical composition which is absent in galaxies without AGN and difficult to carry out in the presence of glare from the direct view of the nucleus seen in quasars.

Studies of chemical composition are likely to be particularly revealing since this is the epoch where we expect the massive host galaxies and the AGN themselves to be in the process of assembly or at least to be in an early stage of their evolution. Since the timescales for AGN phenomena and galactic chemical evolution are very different — by factors of a hundred or so — we can consider the quasar as acting like a 'flash bulb' which gives us a snapshot view of its host galaxy. By studying a number of such objects at a similar redshift, we might expect these snapshots to reveal hosts in different stages of evolution even if the galaxy and AGN formation triggers are closely related.

For objects of this size and complexity, we must be realistic in our expectations for the results of the analysis of emission lines. The volume integrals represented by emission line measurements are taken over a large fraction of the ISM of a whole galaxy and will include inhomogeneities on many scales. The modelling we do vastly oversimplifies the situation but nonetheless may allow us to interpret differences in line ratios from object to object in terms of variations in their properties even if we can place less confidence in, e.g., absolute elemental abundance determinations.

For this study we to use photoionization modelling techniques[23] which include the effects of dust and resonance scattering which are important in this spectral region. The weakness of the Ly$\alpha$ line in the highly polarized sources suggests resonant destruction of the line photons by dust and, indeed, the effect of self-absorption on this line has already been noted in radio galaxies[24]. The anticorrelation between line strength and polarization suggests differences either in the dust/gas ratio or in the spatial distribution of dust from source to source.

The positive correlation between polarization and the NV/CIV ratio suggests that we are witnessing a chemical evolution in the ISM of the galaxies resulting from the star formation activity associated with the assembly of the massive spheroid from smaller, less metal-rich sub-units. Given the small variations in the degree of ionization deduced from the emission line spectra, it is hard to explain the relatively large variations in NV/CIV (almost a dex within the sample) as anything other than a variation in the nitrogen/carbon ratio. A similar effect is seen in the broad line region spectra of quasars[25] and is explained in terms of a galactic evolution model in which the processed material ejected by evolved stars to form galactic winds is trapped by the deep gravitation potential of the spheroid resulting in a rapid pollution of the ISM. The nitrogen enhancement results from secondary nitrogen production in intermediate mass stars over a timescale of around 1 Gyr. Our best-fitting models to the line ratios result from assuming that the nitrogen abundance increases quadratically with the overall metallicity. The fact that we see this effect in the spatially extended, narrow-line regions of these galaxies is important since it indicates that the chemical evolution is happening over the whole galaxy rather than just in the nuclear region sampled by the quasar broad lines. The correlation of this abundance indicator with the degree of continuum polarization is not easy to explain directly but may indicate that the chemical enrichment is accompanied by dust production and dispersal. This would suggest that the end point of the Gyr of rapid chemical evolution is represented by obscured, dusty, metal rich systems which look very much like those of the ultra-luminous infrared galaxies (ULIRG) which are thought to be powered by hidden active nuclei. Comparison of the spectra of gravitationally-lensed ULIRG, e.g., FSC10214+4724[26], with our most polarized sources show that they are closely similar — especially in their NV/CIV, Ly /CIV ratios and degree of polarization.

## 7. CONCLUSIONS

In combination with the detailed imaging provided by HST, the new, large groundbased telescopes have made possible the detailed astrophysical study of the formation and early evolution of massive galaxies. The presence of a massive black hole in these systems is probably ubiquitous and, when active, it can illuminate and ionize a large proportion of the gas in the galaxy. In the particular case of radio galaxies, our direct view to the nucleus is obscured by material which acts as a 'natural coronograph' and frees us from the quasar glare usually affecting observations of their host galaxies. Spectropolarimetry shows us that the restframe ultraviolet continuum of radio galaxies (seen in the optical at these redshifts) is usually dominated by scattered radiation from the obscured nucleus while the narrow emission lines are emitted by the extended interstellar medium and are seen directly. These lines tell the story of the chemical evolution associated with the assembly of what will become the most massive galaxies we know.

The restframe optical spectrum, measured with infrared spectrometers, allows us to detect light from an evolved population of stars and also makes possible the measurement of the familiar forbidden and recombination line spectrum which is so useful for measuring the properties of the gas and reinforcing the conclusion we draw from the ultraviolet lines. The access to the whole spectral range from Ly at 1214Å to H at 6563Å makes it possible for us to study the dominant components which shape the emergent spectrum of these objects. In combination with thermal emission from cool dust and lines from molecules which can be detected by instruments working at longer wavelengths, we can expect to be able to trace the story of the formation of elliptical galaxies in proto-clusters and to follow their evolution to become the giant elliptical galaxies in the cores of today's clusters of galaxies.

## ACKNOWLEDGMENTS

This work is being carried out in collaboration with Marshall Cohen (Caltech), Bob Goodrich (Keck), Joël Vernet & Ilse van Bemmel (ESO), Montse Villar-Martín (Hertfordshire), Sperello di Serego Alighieri & Andrea Cimatti (Arcetri) and Pat McCarthy (OCIW). It is partly based on observations collected at the European Southern Observatory, Paranal, Chile (ESO Programme 63.P-0391). Some of the data presented herein were obtained at the W.M. Keck Observatory, which is operated as a scientific partnership among the California Institute of Technology, the University of California and the National Aeronautics and Space Administration: the Observatory was made possible by the generous financial support of the W.M. Keck Foundation.## REFERENCES